\documentclass[twocolumn,superscriptaddress,secnumarabic,amssymb, nobibnotes,noeprint, aps, prb]{revtex4-2}
\usepackage{CJKutf8}
\usepackage{graphicx,upgreek,amsmath}
\usepackage{hyperref}
\hypersetup{colorlinks=true,linkcolor=blue,citecolor=blue,urlcolor=blue}

\begin{document}
\begin{CJK*}{UTF8}{gbsn} 
\title{Magnetic Bloch-point hopping in multilayer skyrmions and\\
associated emergent electromagnetic signatures}%

\author{Yu Li (李昱)}
\email{YuLi.nano@outlook.com}
\affiliation{Nano Engineering and Spintronic Technologies (NEST) Group, Department of Computer Science, University of Manchester, Manchester M13 9PL, United Kingdom}
\author{Sergiy Mankovsky}
\affiliation{Department Chemie/Physikalische Chemie, Ludwig-Maximilians-Univerit{\"a}t M{\"u}nchen, Butenandtstrasse 5-13, 81377 M{\"u}nchen, Germany}
\author{Svitlana Polesya}
\affiliation{Department Chemie/Physikalische Chemie, Ludwig-Maximilians-Univerit{\"a}t M{\"u}nchen, Butenandtstrasse 5-13, 81377 M{\"u}nchen, Germany}
\author{Hubert Ebert}
\affiliation{Department Chemie/Physikalische Chemie, Ludwig-Maximilians-Univerit{\"a}t M{\"u}nchen, Butenandtstrasse 5-13, 81377 M{\"u}nchen, Germany}
\author{Christoforos Moutafis}
\email{Christoforos.Moutafis@manchester.ac.uk}
\affiliation{Nano Engineering and Spintronic Technologies (NEST) Group, Department of Computer Science, University of Manchester, Manchester M13 9PL, United Kingdom}
\date{June 2021}
\begin{abstract}
    Magnetic multilayers are promising tuneable systems for hosting magnetic skyrmions at/above room temperature.
    Revealing their intriguing switching mechanisms and associated inherent electrical responses are prerequisites for developing skyrmionic devices.
    In this work, we theoretically demonstrate the annihilation of single skyrmions occurring through a multilayer structure, which is mediated by hopping dynamics of topological hedgehog singularities known as Bloch points.
    The emerging intralayer dynamics of Bloch points are dominated by the Dzyaloshinskii-Moriya interaction, and their propagation can give rise to solenoidal emergent electric fields in the vicinity.
    Moreover, as the topology of spin textures can dominate their emergent magnetic properties, we show that the Bloch-point hopping through the multilayer will modulate the associated topological Hall response, with the magnitude proportional to the effective topological charge.
    We also investigate the thermodynamic stability of these states regarding the layer-dependent magnetic properties.
    This study casts light on the emergent electromagnetic signatures of skyrmion-based spintronics, rooted in magnetic-multilayer systems.
\end{abstract}
\maketitle
\end{CJK*}

\section{Introduction}
\begin{figure}[htb]
    \centering
    \includegraphics[width=1\columnwidth]{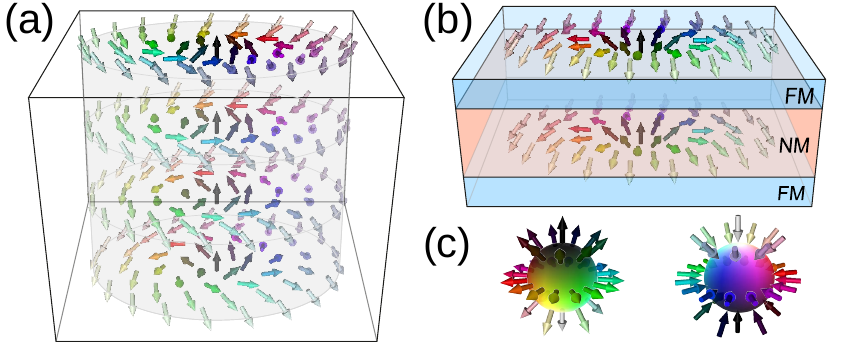}
    \caption{\label{fig:Fig1_Skyrmion}
    Topologically nontrivial spin textures in ferromagnets.
    (a) Bloch-type skyrmion tube in bulk chiral magnets (such as FeGe).
    (b) N{\'e}el-type skyrmion `chain' in magnetic multilayers, with spin configurations only situated in the FM layers.
    (c) Bloch-point singularities with positive (left) and negative (right) topological charges, referred to as BP$^+$ and BP$^-$, respectively.
    }
\end{figure}

Magnetic skyrmions are nontrivial spin textures (Fig.~\ref{fig:Fig1_Skyrmion}) that have remarkable metastability due to their unique topological properties in nature, and have been proposed as candidates of information carriers especially following the recent demonstrations of individual skyrmions in magnetic materials~\cite{Back2020, Fert2017, Nagaosa2019}.
The skyrmion stabilisation highly relies on the competition of isotropic Heisenberg exchange and antisymmetric Dzyaloshinskii-Moriya interaction (DMI)~\cite{Dzyaloshinsky1958, TornMoriya1960}, accompanied by magnetostatic interaction~\cite{Lucassen2019,Li2020}.
Various types of skyrmionic textures have been proposed concerning the class of DMIs arising from different crystal families.
The topological protection property can be characterised by topological charge~\cite{Braun2012, Nagaosa2019}:
\begin{equation}
    Q=\frac{1}{4\pi}\int\mathbf{s}\cdot\left(\partial_x\mathbf{s}\times\partial_y\mathbf{s}\right)\mathrm{d}x\mathrm{d}y\,.
\end{equation}
Importantly, magnetic thin films stacked in nonmagnetic/ferromagnetic (NM/FM) multilayers give rise to a strong spin-orbit coupling, which in combination with the broken symmetry at interfaces can result in a large additive DMI magnitude~\cite{Fert1980, Yang2015,Moreau-Luchaire2016}.
This can enable an excellent tuneability in such material system by precisely engineering the layer properties, making it therefore possible to stabilise magnetic skyrmions even at room temperature~\cite{Moreau-Luchaire2016, Boulle2016, Woo2016}.

The topological constraints of magnetic skyrmions can be broken by the emergence of point singularities in magnets, known as Bloch points~\cite{Malozemoff1979} (Fig.~\ref{fig:Fig1_Skyrmion}c).
They are ultimately small topological defects with magnetisation vanishing in the core, where the continuum micromagnetic approach breaks down.
It has been shown that the Bloch points play crucial roles in mediating the winding/unwinding process of magnetic skyrmions, and their propagation has been thus far proposed in bulk chiral magnets~\cite{Birch2020, Li2020, Milde2013} and magnetic multilayers~\cite{Wohlhuter2015}.
Notably, the discontinuity in the multilayer structures due to the separation of ferromagnetic sites by nonmagnetic spacers, results in much more complex Bloch-point dynamics during the skyrmion unwinding process.
Moreover, the interplay between Bloch points with magnetic skyrmions in three-dimensional multilayers, as well as the associated exotic emergent-field considerations, are particularly intriguing and largely missing.

As one skyrmion holds quantised `emergent flux' upon integration over a unit sphere $S^2$~\cite{Schulz2012,Everschor-Sitte2014,Charilaou2018, Nagaosa2019}, conduction electrons passing through the topologically nontrivial texture will be deflected into the direction transverse to the current flow, causing the topological Hall (TH) effect~\cite{Ye1999,Tatara2007,Neubauer2009,Franz2014,Yin2015,Ndiaye2017,Zeissler2018,Raju2019}.
This type of Hall contribution is invariant and robust against continuous deformation.
The strength of the TH signal, such as TH resistivity $\rho_{xy}^\mathrm{TH}$, is independent of the domain size but purely determined by the topology of the skyrmionic textures.
Therefore, the TH response is able to act as an indicator for the presence of skyrmions, which has been experimentally evidenced in various types of materials, including bulk chiral magnets~\cite{Neubauer2009,Franz2014} and magnetic multilayers~\cite{Raju2019, Zeissler2018}.
It enables a way to read out the skyrmion state by simply measuring the electronic transport response, which has technological relevances for possible future applications of skyrmion-based spintronic devices.
However, the studies so far only focus on the electrical signature of skyrmionic textures in quasi-two-dimensional systems.
Deeper understandings of the signature of skyrmionic textures in magnetic multilayers must necessarily consider $i)$ the coupling of skyrmions across layers~\cite{Legrand2018, Raju2019}, $ii)$ the depth-resolved dynamics of complex spin textures~\cite{Burn2020, Li2020}, and $iii)$ the correlation between the underlying variation of the TH signal and Bloch-point dynamics especially when the integrity of single skyrmions is broken~\cite{Redies2019}, in which case more fundamental studies are essential.

In this work, we start from the chiral magnetisation profile of a single skyrmion in a [Pt/Co/Ta]$_2$ multilayer and identify the complex annihilation dynamics mediated by Bloch points, involving their intralayer propagations as well as the interlayer Bloch-point hopping processes.
The Bloch-point propagation within a Co layer has an oscillatory velocity and gives rise to an emergent electric field with a strikingly high magnitude.
The resulting hopping of Bloch points induces a variation of the TH signal, such as its manifestation in the TH resistivity $\rho_{xy}^\mathrm{TH}$.
Furthermore, we subsequently propose that the intermediate states during the Bloch-point hopping can be stabilised by engineering material properties in a layer-dependent manner, and these states can be well-defined by their TH properties.
Our work then suggests a simple but efficient concept of multi-state skyrmionic devices by fully utilising the emergent electromagnetic signatures embedded in three-dimensional magnetic multilayers.

\section{Bloch-point dynamics in multilayer}
\begin{figure*}[htb]
    \centering
    \includegraphics[width=2\columnwidth]{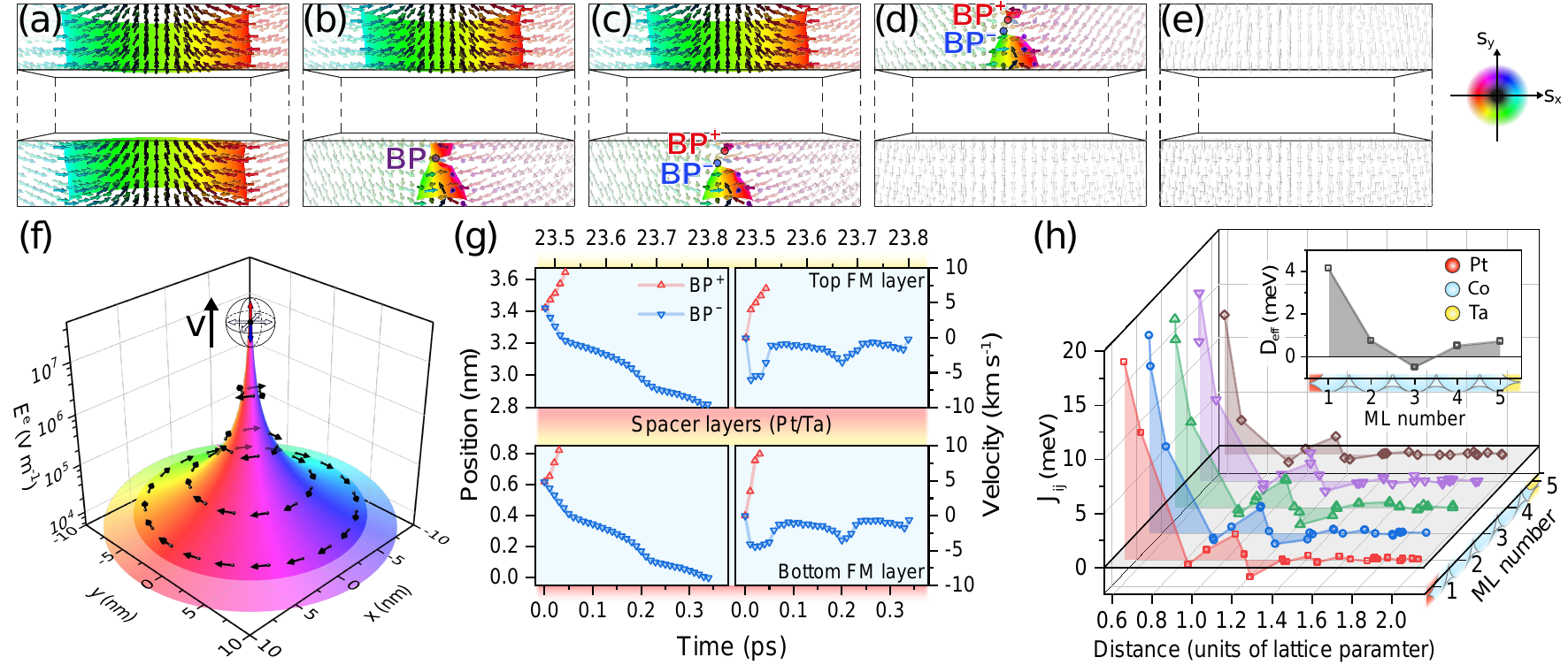}
    \caption{\label{fig:Fig2_BP_intralayer}
    Bloch-point dynamics and associated emergent electric field during the skyrmion annihilation in a [Pt/Co/Ta]$_2$ multilayer.
    (a)--(e) Snapshots of cross-sectional spin profiles in Co layers (Pt and Ta layers are hidden).
    The colour palette represents the spin direction, and isosurfaces indicate the positions of $s_z = 0$.
    (f) Radiation of emergent electric field $\mathbf{E}^\mathrm{e}$ by a moving BP$^+$ with velocity $2\,\mathrm{km\,s^{-1}}$.
    The field has an in-plane vortex-like direction (represented by colour and arrows), with the magnitude illustrated in the $z$ dimension.
    (g) Temporal evolution of the Bloch-point displacement (left-side) and velocity (right-side) along the out-of-plane direction.
    The top and bottom FM layers are represented in blue background, separated by Pt (red) and Ta (yellow) NM layers.
    (h) The pair-wise exchange coupling $J_{ij}$ and layer-resolved effective DMI $D_\mathrm{eff}$ for fcc Co regarding different atomic positions, calculated by the first-principle method.
    }
\end{figure*}

In the multilayer regime, when the thickness of individual FM layers is hitting the limit of micromagnetics, any finer meshes down to sub-nanometre become unrealistic~\cite{Muller2019}.
Therefore, we perform the spin dynamics simulations in an atomistic scheme, using the open-source Spirit code~\cite{Muller2019}.
Here, we consider a [Pt ($1 \,\mathrm{nm}$)/Co/Ta ($1 \,\mathrm{nm}$)]$_2$ system, including $100 \,\mathrm{cell} \times 100 \,\mathrm{cell} \times 5 \,\mathrm{cell}$ in each Co layer with periodic boundary conditions applied in the $xy$ plane.
The pair-wise isotropic exchange interactions $J_{ij}$ (Fig.~\ref{fig:Fig2_BP_intralayer}h) and DMI parameters $\mathbf{D}_{ij}$ are calculated from first-principles (see DFT calculation details in the Note I of the Supplemental Material~\cite{SupplementalMaterial_THE}).
$\mathbf{D}_{ij}$ vectors are used to calculate the effective DMI $D_\mathrm{eff}$ at each monolayer (ML), by analogy with the method in Ref.~\cite{Yang2015} (inset of Fig.~\ref{fig:Fig2_BP_intralayer}h).
Although DMI interactions vanish in bulk fcc Co because of symmetry reasons, the impact of the Co/Pt and Co/Ta interfaces on the electronic structure inside the $5\,\mathrm{MLs}$ Co film is non-vanishing, which leads to finite $D_\mathrm{eff}$ in all the MLs.
Such additive interfacial DMI contributes to the stabilisation of N{\'e}el-type skyrmions in the multilayer structure (Fig.~\ref{fig:Fig2_BP_intralayer}a).
Because the continuity of the ferromagnet is broken by nonmagnetic spacers, the spin texture exists in the form of a skyrmion `chain' (Fig.~\ref{fig:Fig1_Skyrmion}b), and the components hosted at FM layers are coupled by the magnetostatic interaction~\cite{Legrand2018, Legrand2018a}.
Notably, additional interactions between adjacent FM layers may also contribute to the skyrmion stabilisation in magnetic multilayers, such as Ruderman-Kittel-Kasuya-Yoshida (RKKY) effect~\cite{Parkin1991,Legrand2020}.
However, regarding the selection of materials in this work, the RKKY coupling of Pt or Ta is sufficiently weak~\cite{Parkin1991}, making it reasonable to assume that the interlayer coupling is dominated by the magnetostatic interaction~\cite{Lucassen2019}.
Such stabilisation mechanism is different from that of single skyrmion tubes in bulk chiral magnets~\cite{Li2020, Milde2013} (Fig.~\ref{fig:Fig1_Skyrmion}a), where both exchange and magnetostatic couplings coexist throughout the sample.
The strength of the long-range magnetostatic coupling is also expected to be smaller than the Heisenberg exchange.
Thus, the skyrmion annihilation at the critical field (see Video 1 in the Supplemental Material~\cite{SupplementalMaterial_THE}) is not initiated simultaneously in all FM layers but starts to unwind from the bottom, accompanied by the skyrmion deformation into two opposing chiral bobbers with a pair of Bloch points, and each corresponds to a `bobber' (Fig.~\ref{fig:Fig2_BP_intralayer}b).
It is worth noting that the Bloch-point pair is created near the middle of the FM layer, which can be explained by the spatially inhomogeneous $D_\mathrm{eff}$ (inset of Fig.~\ref{fig:Fig2_BP_intralayer}h), i.e., the DMI near the Pt/Co interface (ML number $= 1$) holds much larger magnitude than that near the Co/Ta interface (ML number $= 5$), which provides more robust stability; moreover, the $D_\mathrm{eff}$ in the middle of a Co layer (ML number $= 3$) is the weakest and even has an opposite sign to those near the two interfaces, having an effect on destabilising the skyrmion in the FM layer.
After being created, the two Bloch points (BP$^+$ and BP$^-$) propagate upward and downward, respectively, until finally disappear at the FM/NM interfaces (Fig.~\ref{fig:Fig2_BP_intralayer}c).
Then affected by the demagnetising field of the lower FM layer, a new Bloch-point pair is created at the upper FM layer, with the subsequent dynamics following a similar manner (Fig.~\ref{fig:Fig2_BP_intralayer}d--e).

The created Bloch points act as magnetic monopoles/anti-monopoles, and produce radial emergent magnetic fields $\mathbf{B}^\mathrm{e} = g^\mathrm{e}\mathbf{r}/r^3$ $(r > 0)$ with quantised emergent charge $g^\mathrm{e}=\pm\hbar/2q^\mathrm{e}$ ($q^\mathrm{e} = \pm1/2$ for majority/minority electron spins)~\cite{Griffiths2017, Seidel2016}.
In the adiabatic limit, for a rigid magnetic Bloch-point texture moving with a drifting velocity $\mathbf{v}_\mathrm{BP}$, its magnetisation $\mathbf{m}$ can be considered only as a function of time $\mathbf{m}(\mathbf{r},t) = \mathbf{m}(\mathbf{r} - \mathbf{v}_\mathrm{BP}t)$.
By analogy with the Faraday's law of induction by $\mathbf{E}^\mathrm{e} = -\mathbf{v}_\mathrm{BP}\times \mathbf{B}^\mathrm{e}$~\cite{Charilaou2018, Li2020, Schulz2012}, the propagation velocity of a Bloch point (e.g.~BP$^+$) has a magnitude in the order of kilometres per second, and induces an in-plane real electric field in a vortex form perpendicular to the radial direction with a magnitude in the order of megavolts per metre (Fig.~\ref{fig:Fig2_BP_intralayer}f).
Although these two Bloch points propagate in opposite directions (Fig.~\ref{fig:Fig2_BP_intralayer}g), their topological charges are also opposite, and as a result, the superimposed electric field radiates in the same direction.
It should be emphasised that the propagation is periodically modulated by discrete potential wells between neighbouring fcc Co lattice sites (Fig.~\ref{fig:Fig2_BP_intralayer}g)~\cite{Kim2013, Li2020} as well as the artificial pinning imposed by the interlayer spacer (Fig.~S1 in the Supplemental Material~\cite{SupplementalMaterial_THE}) which can be experimentally tuned.
The velocity magnitude of the Bloch point moving downwards (BP$^-$) oscillates and reaches another peak at ${\sim}0.2 \,\mathrm{ps}$ (${\sim}23.5 \,\mathrm{ps}$) in the bottom (top) FM layer (bottom-right or top-right panel of Fig.~\ref{fig:Fig2_BP_intralayer}g), with the frequency lying in a terahertz range.
It should be noted that its dynamic behaviour as well as the associated emergent electric field can be manipulated by varying magnetic parameters~\cite{Li2020}, which should be fundamentally relevant to the phase velocity of the spontaneously emitting spin wave and the propagation velocity of the Bloch point~\cite{Bouzidi1990,Hertel2016,Ma2020}.

\section{Interlayer Bloch-point hopping and topological Hall signal}
\begin{figure*}[htb]
    \centering
    \includegraphics[width=2\columnwidth]{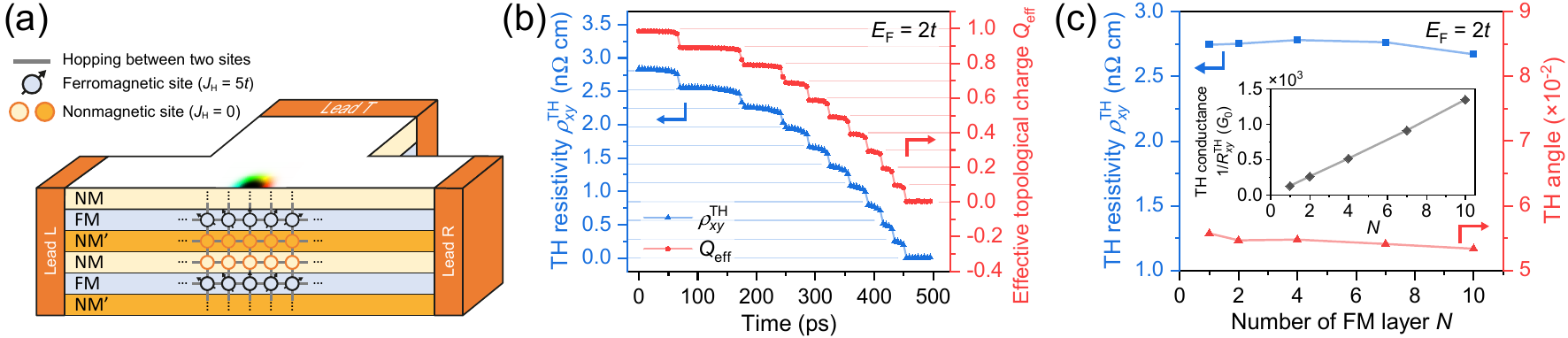}
    \caption{\label{fig:Fig3_BP_interlayer}
    Topological Hall effect of a multilayer skyrmion.
    (a) Schematic of a multilayer [NM/FM/NM']$_N$ transport model.
    The cross-section of a four-terminal set-up is shown, including a spin texture placed in the centre.
    (b) Temporal evolution of $\rho_{xy}^\mathrm{TH}$ (blue triangles) and effective topological charge $Q_\mathrm{eff} = \frac{1}{N}\sum Q_n$ (red circles) during the annihilation/unwinding process.
    (c) TH responses including TH resistivity $\rho_{xy}^\mathrm{TH}$ (blue squares), TH angle $\theta^\mathrm{TH}$ (red triangles) and TH conductance $1/R_{xy}^\mathrm{TH}$ (black diamonds in the inset) regarding the total thickness of FM layers (number of FM layers $N$).
    Solid lines connecting data points are guides to the eye.
    }
\end{figure*}

It is known that magnetic skyrmions can provide quantised topological Hall (TH) response for transport measurements, which is independent of the skyrmion diameter or applied magnetic field~\cite{Ndiaye2017, Zeissler2018}.
It thus points to a superior solution to track the anomalous Hall effect~\cite{Nagaosa2010} for skyrmion characterisation by the unambiguous electrical detection that is linked to potential applications.
Previous work~\cite{Ndiaye2017, Zeissler2018} has shown the intrinsic connection of the TH signal and the number of skyrmions by utilising their topological nature.
However, studies on the emergent TH response properties are necessary when the integrity of skyrmionic textures is broken~\cite{Redies2019}.
As we have shown, the unwinding of single skyrmions in magnetic multilayers evolves in an inhomogeneous manner as a function of depth.
To elucidate the inherent TH signature of such three-dimensional textures, we study the variation of TH resistivity $\rho_{xy}^\mathrm{TH}$ during the skyrmion annihilation in a [NM/FM/NM']$_{10}$ multilayer.
The magnetisation dynamics is simulated by a micromagnetic model using mumax$^3$~\cite{Vansteenkiste2014} (see results in the Note II of the Supplemental Material~\cite{SupplementalMaterial_THE}), as it is more efficient than atomistic models when computing a large-scale system, especially on the calculation of demagnetising fields.
We then numerically verify the transport properties by following a single-orbital tight-binding model~\cite{Yin2015, Ndiaye2017,Gobel2019} using the Kwant package~\cite{Groth2014} (schematic of the crossbar set-up shown in Fig.~\ref{fig:Fig3_BP_interlayer}a),
where the interaction of electrons with local magnetisation fields is described by~\cite{Nagaosa2010, Ndiaye2017, Yin2015}:
\begin{equation}
    H_\mathrm{e}=\sum_{i}{c_i^\dagger\epsilon_ic_i}-t\sum_{\left\langle i,j\right\rangle}\left(c_i^\dagger c_i + \mathrm{H.c.}\right) -J_\mathrm{H} \sum_i c_i^\dagger\boldsymbol{\upsigma} c_i\cdot \mathbf{m}_i,
\end{equation}
where $J_\mathrm{H}$ is the Hund's coupling between the spin $\boldsymbol{\upsigma}$ of itinerant electrons and the magnetisation $\mathbf{m}_i$.
$\epsilon_i$ is the on-site energy.
$t$ is the hopping integral between nearest-neighbour sites $i$ and $j$, determined by the effective mass of electrons $m^\ast$ and distance of sites $a$: $t=\hbar^2/2m^\ast a^2$.
$c_i^\dagger (c_i)$ is the creation (annihilation) operator of site $i$.
Electronic transmissions are calculated by the Landauer-B{\"u}ttiker formalism~\cite{Datta1995} (see method in the Note III of the Supplemental Material~\cite{SupplementalMaterial_THE}).

We consider a defect-free system without impurity scatterings, and the on-site energy is neglected.
We also assume a strong coupling with $J_\mathrm{H} = 5t$ based on the adiabatic approximation~\cite{Yin2015}, where an electron spin will be constantly adjusted and aligned to the magnetisation direction when passing through a magnetic texture.
The annihilation process consists of gradually unzipping the skyrmion `chain' from the bottom-most FM layer towards the top (Fig.~S1 in the Supplemental Material~\cite{SupplementalMaterial_THE}).
At the beginning, an intact skyrmion `chain' has the effective topological charge $Q_\mathrm{eff} \approx 1$ (Fig.~\ref{fig:Fig3_BP_interlayer}b).
As it gradually unwinds, breaking into its components, $Q_\mathrm{eff}$ experiences a stairlike decrease until reaches `0' when the skyrmion `chain' is fully unwound to a spin-polarised state.
By plotting the temporal evolution of TH resistivity $\rho_{xy}^\mathrm{TH}$ (Fig.~\ref{fig:Fig3_BP_interlayer}b), we find that the magnitude is approximately proportional to $Q_\mathrm{eff}$.

To further elaborate the intrinsic connection of the TH effect with the effective topological charge of a single skyrmion `chain', we focus on the TH response of the magnetic multilayer [NM/FM/NM']$_N$ regarding the total thickness of FM layers (expressed as the number of repeats $N$).
As each component of the `chain' provides the same amount of channels for electron deflection, the total TH conductance ($1/R_{xy}^\mathrm{TH}$) is proportional to $N$ (inset of Fig.~\ref{fig:Fig3_BP_interlayer}c), whereas both TH resistivity $\rho_{xy}^\mathrm{TH}$ and TH angle $\theta^\mathrm{TH}$ ($V_\mathrm{H}/V_x$) are independent of $N$ (Fig.~\ref{fig:Fig3_BP_interlayer}c).
These findings suggest that $\rho_{xy}^\mathrm{TH}$ can be considered as the `density' of the electron deflection contributed by the $N$ FM layers of the skyrmion `chain'.
Notably, the variation of the transport properties in a bulk chiral system is reported by Redies et al.~\cite{Redies2019}, where the Hall conductivity decreases as the heights of chiral bobbers increase.
Furthermore, the associated $\mathbf{B}^\mathrm{e}$ of Bloch points will have a pronounced influence on the electronic states.
Their demonstration~\cite{Redies2019} is similar to our findings shown in Fig.~\ref{fig:Fig3_BP_interlayer}b, where a drastic decrease of the TH resistivity $\rho_{xy}^\mathrm{TH}$ occurs with the presence of Bloch points during the breaking of a skyrmion `chain'.
In previous work~\cite{Neubauer2009, Raju2019, Tatara2007, Zeissler2018}, $\rho_{xy}^\mathrm{TH}$ is associated with $\rho_{xy}^\mathrm{TH} = PR_0 B_\mathrm{eff}^z = PR_0\phi_0q_\mathrm{d}$ in skyrmionic systems, with the spin polarisation $P$, ordinary Hall coefficient $R_0$, and flux quantum $\phi_0$.
$q_\mathrm{d} = Q/A$ is the density of 2D topological charge $Q$ within an area $A$~\cite{Raju2019, Zeissler2018}.
Herein, we extend this equation by considering the overall 3D topology in terms of non-uniform topological charge density $q_{\mathrm{d},n}$ in the $n^\mathrm{th}$ FM layer:
\begin{equation}
    \rho_{xy}^\mathrm{TH} = P R_0 \phi_0 \frac{1}{N}\sum_{n=1}^N q_{\mathrm{d},n}\,.
\end{equation}
Conceptually, the variation of $\rho_{xy}^\mathrm{TH}$ can be attributed to the topological winding/unwinding, which fundamentally manifests the Bloch-point hopping across a distance $\Delta x_\mathrm{BP}$:
\begin{equation}
    \Delta \rho_{xy}^\mathrm{TH} = P R_0 \phi_0 \frac{\Delta Q}{A} = P R_0 \phi_0 \frac{ \Delta x_\mathrm{BP}}{V}\,.
\end{equation}
Therefore, the dynamics of infinitely small Bloch points within the three-dimensional multilayers overall volume $V$ can be indirectly connected.
As this phenomenon is also simultaneously accompanied by radiated emergent electric fields $\mathbf{E}^\mathrm{e}$, these results may therefore provide essential ramifications for topological spin textures in nanoscale magnetic multilayers and the link to their electrical detections and potential applications exploiting their associated emergent electromagnetic signatures.

\begin{figure*}[t]
  \centering
  \includegraphics{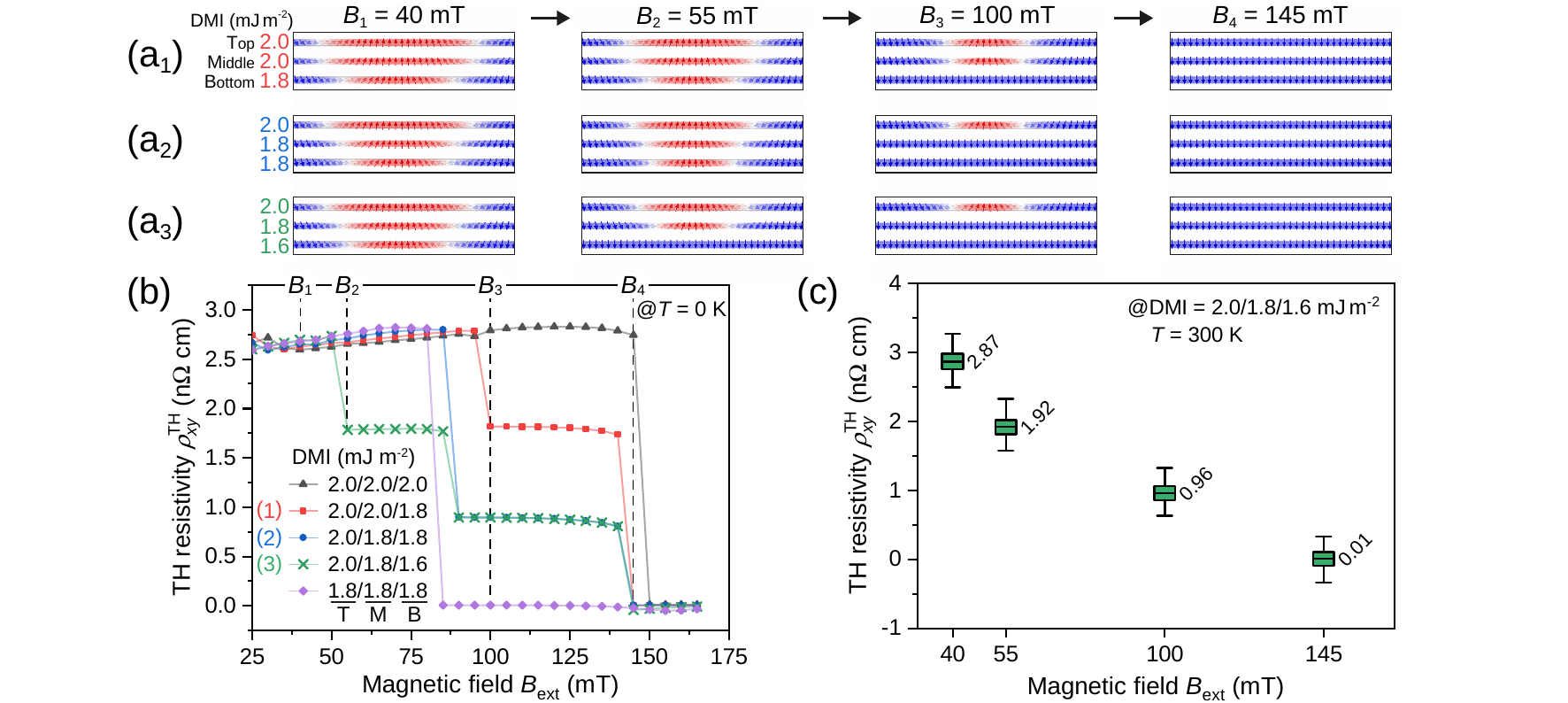}
  \caption{\label{fig:Fig4_Eng_DMI}
  Multiple metastable states with distinct TH signatures by engineering layer-dependent properties in [NM/FM/NM']$_3$ multilayers.
  (a) Annihilation processes of a single skyrmion accompanied by multi-state behaviours, regarding three cases (a$_1$)--(a$_3$) of layer-dependent DMIs.
  (b) Variation of $\rho_{xy}^\mathrm{TH}$ as a function of $B_\mathrm{ext}$, including the cases corresponding to (a$_1$) with red squares, (a$_2$) with blue circles, and (a$_3$) with green crosses.
  (c) Statistical means and deviations of $\rho_{xy}^\mathrm{TH}$ against thermal fluctuations, regarding (a$_3$) with $\mathrm{DMI} = 2.0/1.8/1.6\,\mathrm{mJ\,m^{-2}}$ at the T(op)/M(iddle)/B(ottom) FM layer.
  }
\end{figure*}

\section{Engineering layer-dependent properties for 3D multistates}
To take a step further, we propose a prototype for three-dimensional skyrmionic memory and logic devices by utilising metastable states of skyrmion `chains' in magnetic multilayers, which would be relevant to room-temperature multi-bit technological solutions.
Previous work~\cite{Gabor2017, Khadka2018, Legrand2020} demonstrated various routes of tailoring the magnetic properties (e.g.~anisotropy, DMI and RKKY coupling) by directly engineering FM layers or FM/NM interfaces.
Here, we exemplify the [NM/FM/NM']$_3$ systems where the DMIs are engineered in a layer-dependent manner (Fig.~\ref{fig:Fig4_Eng_DMI}a).
Taking an example of Fig.~\ref{fig:Fig4_Eng_DMI}a$_3$, where the DMI constant is $2.0/1.8/1.6\,\mathrm{mJ\,m^{-2}}$ at the top/middle/bottom FM layer, the energy barrier for the Bloch-point creation can be sufficiently lowered in a FM layer with smaller DMI.
As a result, the annihilation processes occur at sufficiently distinct magnetic fields $B_\mathrm{ext} = 140/90/50\,\mathrm{mT}$ (see method of `field sweeping' in the Note II of the Supplemental Material~\cite{SupplementalMaterial_THE}), and the resulting TH resistivity $\rho_{xy}^\mathrm{TH}$ also shows a multi-level feature as a function of the field (green crosses in Fig.~\ref{fig:Fig4_Eng_DMI}b).
Furthermore, to test the stability of the metastable states (shown in Fig.~\ref{fig:Fig4_Eng_DMI}a$_3$) against finite temperature, each state starts from its zero-kelvin equilibrium (in Fig.~\ref{fig:Fig4_Eng_DMI}b), and a randomly fluctuating thermal field~\cite{Brown1963} ($T = 300\,\mathrm{K}$) is applied for $20\,\mathrm{ns}$.
During the process $\rho_{xy}^\mathrm{TH}$ is recorded every $0.05\,\mathrm{ns}$ for each field and the statistical mean and deviation are calculated.
In Fig.~\ref{fig:Fig4_Eng_DMI}c, these states can be well-defined, which are approximately proportional to the length of the remnant `chain'.
We should emphasise that the dynamics module in micromagnetics may not precisely emulate the realistic thermal effect for probing the thermostability of spin textures, so future dedicated studies would be needed systematically to evaluate the lifetime of metastable states embedded in magnetic multilayers.

Multi-level bits defined by layer-dependent magnetisation states have been demonstrated in various prototypes, including magnetic ratchets~\cite{Lavrijsen2013} and magnetic abacus memories~\cite{Zhang2014}.
The device-level performance is expected to be more reliable against external perturbations by utilising the TH signals, because the topological stability effectively protects the topological charge from continuous variations.
Furthermore, in realistic conditions, disordered pinning sites which generally exist in sputtered films may widen the range of critical fields for skyrmion annihilation, or may energetically favour worm-like textures~\cite{Raju2019}.
In such cases the system undergoes gradual metamagnetic transitions which are reflected in a smooth variation of $\rho_{xy}^\mathrm{TH}-B_\mathrm{ext}$ loops~\cite{Pierobon2018, Raju2019, Soumyanarayanan2017} rather than the sharp jumps shown in Fig.~\ref{fig:Fig4_Eng_DMI}b.
The layer-dependent discontinuities and inhomogeneities of magnetic properties here act as new types of `soft pinning' and crucially alter the energy barriers among the metastable states, making it possible to energetically compete against intralayer disorders and even dominate the state transitions.

\section{Conclusions}
In conclusion, we have investigated single skyrmions in magnetic multilayers that can exist as skyrmion `chains', hosted in FM layers and indirectly coupled by magnetostatic interactions.
By the application of an external out-of-plane magnetic field, a skyrmion `chain' starts to unwind from the bottom FM layer, and the process is mediated by complex dynamics of Bloch points: a pair of Bloch points is created from the weakest-DMI part of an FM layer, and the two Bloch points move towards opposite directions until they finally disappear at the strongest-DMI interfaces.
Similar to the mechanism in bulk chiral magnets~\cite{Li2020}, the Bloch-point propagation gives rise to a vortex-like emergent electric field perpendicular to the propagation direction.
The magnitude is in the order of megavolts per metre, and its oscillation frequency lies in the terahertz range.
On the other hand, when the Bloch point hops through the multilayer, its displacement induces a variation in the TH response, and notably, its resistivity $\rho_{xy}^\mathrm{TH}$ experiences a stairlike decrease with the magnitude approximately proportional to the effective topological charge.
Furthermore, we propose that metastable states in the skyrmion `chain' can be thermodynamically stabilised by tailoring the layer-dependent properties of the materials, by means of, for example, engineering different DMI in the layers.
These states have distinct TH signatures, and can be switched by simply varying the external magnetic field.
Our findings elucidate magnetic multilayer skyrmions and their three-dimensional topology and could significantly enrich the availability of transport signatures in multilayer nontrivial spin textures.
In addition, this work proposes a paradigm of three-dimensional devices embedded with multi-bit functionalities.
By building on recent proposals in three-dimensional spintronics~\cite{Fernandez-Pacheco2017, Lavrijsen2013, Zhang2014}, our results point to fruitful directions of practical room-temperature applications, in technologically relevant multilayers where experimental evidence is within reach.

\begin{acknowledgments}
The authors would like to acknowledge the assistance given by Research IT and the use of the Computational Shared Facility at The University of Manchester.
Financial support by the DFG via SFB 1277 (Emergente Relativistische Effekte in der Kondensierten Materie) is gratefully acknowledged.
Y.L. acknowledges the funding supports by The University of Manchester and the China Scholarship Council.
\end{acknowledgments}

%
  
\end{document}


\begin{CJK*}{UTF8}{gbsn} 
\title{Supplemental material for\\
``Magnetic Bloch-point hopping in multilayer skyrmions and \\
associated emergent electromagnetic signatures''}%

\author{Yu Li (李昱)}
\email{YuLi.nano@outlook.com}
\affiliation{Nano Engineering and Spintronic Technologies (NEST) Group, Department of Computer Science, University of Manchester, Manchester M13 9PL, United Kingdom}
\author{Sergiy Mankovsky}
\affiliation{Department Chemie/Physikalische Chemie, Ludwig-Maximilians-Univerit{\"a}t M{\"u}nchen, Butenandtstrasse 5-13, 81377 M{\"u}nchen, Germany}
\author{Svitlana Polesya}
\affiliation{Department Chemie/Physikalische Chemie, Ludwig-Maximilians-Univerit{\"a}t M{\"u}nchen, Butenandtstrasse 5-13, 81377 M{\"u}nchen, Germany}
\author{\\Hubert Ebert}
\affiliation{Department Chemie/Physikalische Chemie, Ludwig-Maximilians-Univerit{\"a}t M{\"u}nchen, Butenandtstrasse 5-13, 81377 M{\"u}nchen, Germany}
\author{Christoforos Moutafis}
\email{Christoforos.Moutafis@manchester.ac.uk}
\affiliation{Nano Engineering and Spintronic Technologies (NEST) Group, Department of Computer Science, University of Manchester, Manchester M13 9PL, United Kingdom}
\maketitle
\end{CJK*}

\begin{flushleft}
\textbf{Table of contents}

Video 1: Atomistic result of the annihilation of a single skyrmion `chain' in a [Pt/Co/Ta]$_2$ multilayer.

Video 2: Micromagnetic result of the annihilation of a single skyrmion `chain' in a [NM/FM/NM']$_{10}$ multilayer.
\\~\\
Supplementary notes

\hyperref[sec:Note_I]{Note I.} DFT calculations: details and some results.

\hyperref[sec:Note_II]{Note II.} Annihilation of single multilayer skyrmion implemented by micromagnetic simulations.

\hyperref[sec:Note_III]{Note III.} Landauer-B\"uttiker formalism for charge currents.

\end{flushleft}
\newpage

\section{DFT calculations: details and some results}  \label{sec:Note_I}
The interatomic exchange coupling parameters, $J_{ij}$ and $\mathbf{D}_{ij}$, are calculated within the multiple-scattering formalism~\cite{Ebert2009,Mankovsky2017}, based on first-principles electronic structure calculated using the spin-polarised relativistic KKR (SPR-KKR) Green function method~\cite{Ebert2011,Ebert2017} performed within the framework of the local spin-density approximation (LSDA) to density functional theory (DFT), using a parametrisation for the exchange and correlation potential as given by Vosko et al.~\cite{Vosko1980}.
For the angular momentum expansion of the Green function a cutoff of $l_\mathrm{max} = 3$ was applied.
A ($69 \times 69 \times 4$) $k$-point grid was used for the integration over the Brillouin zone.

The electronic structure calculations for the Pt/Co/Ta multilayer have been performed using a system with a periodic supercell composed of $5 \,\mathrm{ML}$ Pt, $5 \,\mathrm{ML}$ Co and $5 \,\mathrm{ML}$ Ta.
As there is no evidence for the ideal crystalline structure in this multilayer system studied experimentally~\cite{Bai2021}, within present calculations the structure of the system was approximated by the (111)-oriented fcc layers with the lattice parameter $a = 3.92 \,\textrm{\AA}$ corresponding to the Pt fcc lattice.
This would imply the multilayer system is grown epitaxially on the Pt(111) substrate, exhibiting structural relaxation due to modification of the interatomic distances along the $z$ direction across the unit cell.
Auxiliary calculations have been performed using the VASP package~\cite{Kresse1993,Kresse1994} in order to determine the optimised structure parameters.
In these calculations, the generalised gradient approximation (GGA) density functional for the exchange and correlation potential with the Perdew-Burke-Ernzerhof (PBE) parametrisation scheme has been used, as given by Perdew et al.~\cite{Perdew1996}.
A plane wave basis set up to a cutoff energy of $440 \,\mathrm{eV}$ was used for the wave function representation.

The structure optimisation results in ${\sim}18\%$ decrease of the interlayer distance in the Co film when compared to that corresponding to an ideal fcc lattice, while in the Ta film the interlayer distance increases by ${\sim}16\%$.
The Co--Pt and Co--Ta distances at the interface are decreased by ${\sim}16\%$ and ${\sim}4\%$, respectively.
The electronic structure calculations using the SPR-KKR package~\cite{Ebert2011} lead to spin magnetic moments in Co film as $1.712, 1.639, 1.638, 1.673, 1.102\,\mu_\mathrm{B}$, with the smallest value corresponding to the Co/Ta interface.

\section{Annihilation of single multilayer skyrmion implemented by micromagnetic simulations} \label{sec:Note_II}
\begin{figure*}[htb]
    \centering
    \includegraphics[width=1\columnwidth]{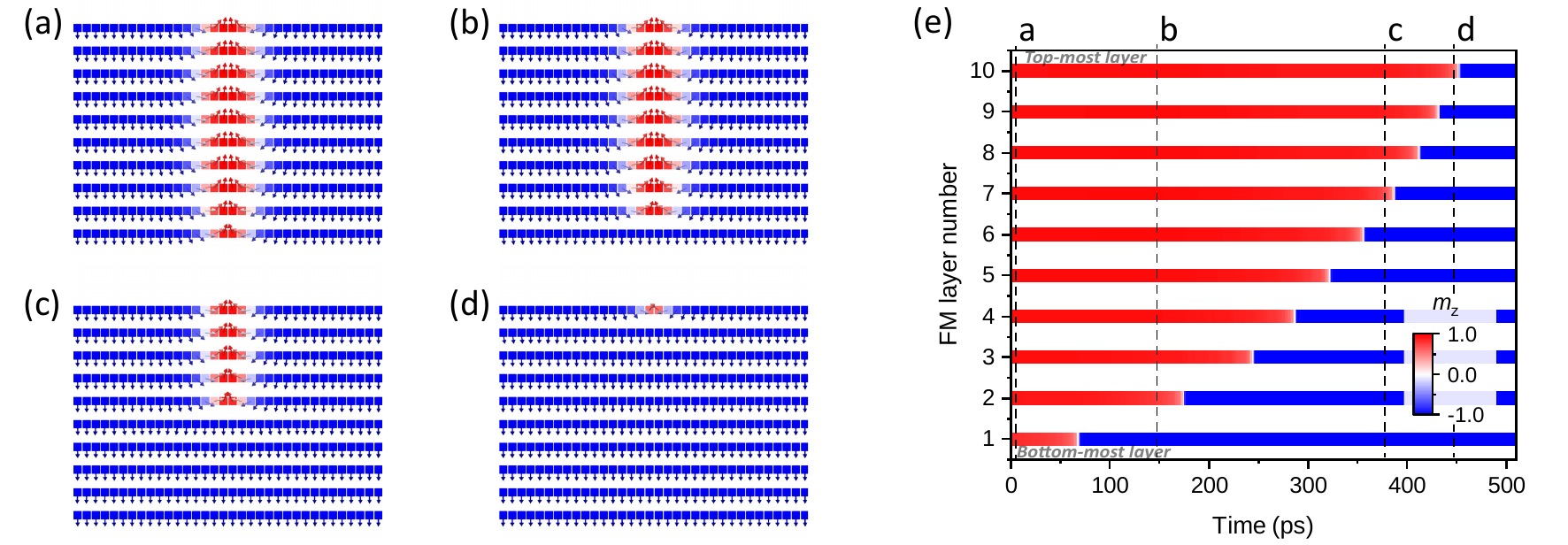}
    \caption{\label{fig:FigS1_Anni_micro}
    Time-dependent annihilation process of a multilayer skyrmion with $10$ FM layers.
    (a)--(d) Snapshots of cross-sectional $m_z$ profiles.
    (e) Temporal evolution of $m_z$ in the skyrmion centre.
    NM layers are shown as white intervals.
    }
\end{figure*}

Here we consider the annihilation process of a single skyrmion in a [NM (1nm)/FM (1nm)/NM' (1nm)]$_{10}$ multilayer crossbar, with the length of $200\,\mathrm{nm}$ and the effective area of $100\,\mathrm{nm} \times 100\,\mathrm{nm}$.
The system is discretised into cuboid cells with the size of $2\,\mathrm{nm}\times 2\,\mathrm{nm}\times 1\,\mathrm{nm}$.
The simulation is performed by a micromagnetic model using mumax$^3$~\cite{Vansteenkiste2014}, and the typical materials parameters used in the simulations are~\cite{Moreau-Luchaire2016}: the Gilbert damping $\alpha = 0.01$; saturation magnetisation $M_\mathrm{s} = 956\,\mathrm{kA\,m^{-1}}$; exchange stiffness $A_\mathrm{ex} = 10\,\mathrm{pJ\,m^{-1}}$; interfacial DMI constant $D_\mathrm{int} = 2\,\mathrm{mJ\,m^{-2}}$; uniaxial perpendicular magnetic anisotropy $K_\mathrm{u1} = 717\,\mathrm{kJ\,m^{-3}}$.
A N{\'e}el-type skyrmion can be stabilised in the form of a skyrmion `chain' hosted in FM layers and indirectly coupled together by magnetostatic interaction (Fig.~\ref{fig:FigS1_Anni_micro}a).

Then a magnetic field of opposite direction to the core polarity is applied, whose magnitude gradually increases with a step of $5\,\mathrm{mT}$ and a duration of $5\,\mathrm{ns}$ at each step.
The energy convergence is monitored during the dynamic `field sweeping' to confirm the equilibrium states.
The process initiates the decreasing of the skyrmion diameter, and at a critical field, the skyrmion `chain' starts to unwind from the bottom FM layer, followed by continuous unwindings in one FM layer after another, all throughout the multilayer structure (Fig.~\ref{fig:FigS1_Anni_micro}a--d), and see the dynamic process in Video 2).
The dramatic annihilation process occurs within $500\,\mathrm{ps}$ (Fig.~\ref{fig:FigS1_Anni_micro}e).
Note that the site of the skyrmion unwinding (from the bottom layer) is determined by the collective effect of the sign of DMI constant and the magnetostatic interaction~\cite{Li2020}.

\section{Landauer-B\"uttiker formalism for charge currents} \label{sec:Note_III}
In the four-terminal set-up (Fig.~3a in the main text), the current $I_m$ flowing through the lead $m$ is the sum of transmission coefficient $T_{mn}$ times voltage $V_n$ at the lead $n$~\cite{Datta1995,Yin2015,Ndiaye2017}:
\begin{equation}
I_m = \frac{e^2}{h} \sum_n T_{mn}V_n\,,
\end{equation}
with $m,n \in \{\mathrm{L,T,R,B}\}$.
The calculation of $T_{mn}$ is based on the Green's function, where $T_{mn}=\mathrm{Tr}(\Gamma_m \mathrm{G}_{mn}^\mathrm{R} \Gamma_n \mathrm{G}_{mn}^{\mathrm{R}\dagger})$, with $\Gamma_m = i (\Sigma_m - \Sigma_m^\dagger)$ and the retarded Green's function $\mathrm{G}^\mathrm{R} = [E-H_\mathrm{e}-\sum \Sigma_m]^{-1}$.
$\Sigma_m = t^2 [\mathrm{G}_m^\mathrm{R} ]^{-1}$ is the self-energy of the lead $m$.

As we are only interested in the final steady state where the charge current must be conserved, one of the terminal voltages can be defined to zero without losing the generality ($V_\mathrm{B} = 0$ in this case).
Then in the simulation of Hall measurement, we have a current $I$ flowing along the longitudinal direction $I_\mathrm{L} = -I_\mathrm{R} = I$ and zero transverse current $I_\mathrm{T} = I_\mathrm{B} = 0$.
As a result, we obtain a reduced invertible matrix:
\begin{equation}
  \left[I,0,-I\right]^\mathrm{T} = \frac{e^2}{h} \mathcal{T} \left[V_\mathrm{L},V_\mathrm{T},V_\mathrm{R}\right]^\mathrm{T}\,,
\end{equation}
where the superscript `T' indicates the transpose of the matrix, and the transmission matrix is given by $\mathcal{T}$.
Then the terminal voltages can be expressed as:
\begin{subequations}
  \begin{align}
      V_\mathrm{L} &= \frac{h}{e^2}\left[\mathcal{R}_{11}-\mathcal{R}_{13}\right] I\,,\\
      V_\mathrm{T} &= \frac{h}{e^2}\left[\mathcal{R}_{21}-\mathcal{R}_{23}\right] I\,,\\
      V_\mathrm{R} &= \frac{h}{e^2}\left[\mathcal{R}_{31}-\mathcal{R}_{33}\right] I\,,\\
      V_\mathrm{B} &= 0\,,
  \end{align}
\end{subequations}
with $\mathcal{R} = \mathcal{T}^{-1}$.
These further allow to calculate the topological Hall (TH) signals:
\begin{subequations}
  \begin{align}
  &\textrm{TH angle: }
  \theta^\mathrm{TH} = \arctan\left(\frac{V_\mathrm{T} - V_\mathrm{B}}{V_\mathrm{L} - V_\mathrm{R}}\right) = \arctan\left(\frac{\mathcal{R}_{21}-\mathcal{R}_{23}}{\mathcal{R}_{11} + \mathcal{R}_{33} - \mathcal{R}_{13} - \mathcal{R}_{31}}\right)\,,\\
  &\textrm{TH resistance: }
  R_{xy}^\mathrm{TH} = \frac{V_\mathrm{T} - V_\mathrm{B}}{I} = \frac{h}{e^2}\left[\mathcal{R}_{21} - \mathcal{R}_{23}\right]\,,\\
  &\textrm{TH resistivity: }
  \rho_{xy}^\mathrm{TH} = \frac{R_{xy}^\mathrm{TH}}{\mathrm{Thickness}}\,.
  \end{align}
\end{subequations}

%
  